\documentclass[10pt,twoside]{epnt1p}
\usepackage{epsfig}
\usepackage{xspace}   

\usepackage{amssymb}
\usepackage{amsmath}

\usepackage{url}

\newcommand{\DS}[1]{$\mathsf{#1}$\xspace}
\newcommand{\fracnew}[2]
           {\protect\frac{{#1}_{\protect\vphantom{!_a}}}{{#2}^{\protect\vphantom{a}}}}

\newcommand{\bdi}{\begin{displaymath}}
\newcommand{\edi}{\end{displaymath}}
\newcommand{\bfi}{\begin{figure}}
\newcommand{\efi}{\end{figure}}

\newcommand{\beq}{\begin{equation}}
\newcommand{\eeq}{\end{equation}}
\newcommand{\beqa}{\begin{eqnarray}}
\newcommand{\eeqa}{\end{eqnarray}}

\newcommand {\BN}     {Bogoliubov--Nambu}

\newcommand {\SM}     {Standard Model}
\newcommand {\gsim}{\mathrel{\hbox{\rlap{\lower.55ex \hbox {$\sim$}}
            \kern-.3em \raise.4ex \hbox{$>$}}}}
\newcommand {\lsim}{\mathrel{\hbox{\rlap{\lower.55ex \hbox {$\sim$}}
            \kern-.3em \raise.4ex \hbox{$<$}}}}
\begin{document}

\begin{frontmatter}


\title{Emergent Relativity: Neutrinos as Probe of the Underlying Theory}


\author[address1]{F.R. Klinkhamer}

\address[address1]{Institute for Theoretical Physics,
             University of Karlsruhe (TH), 76128 Karlsruhe, Germany}

\begin{abstract}
Neutrinos allow for a test of the hypothesis that
the fermions of the Standard Model have Fermi-point splitting,
analogous to the fermionic quasi-particles of
certain condensed-matter systems. If present,
the corresponding Lorentz-violating terms in the Hamiltonian
may provide a new source
of \DS{T} and \DS{CP} violation in the leptonic sector, which is
not directly related to mass.
\end{abstract}



\end{frontmatter}


\section{\label{FRK-sec:intro} Introduction}

The basic idea of this talk is to suggest
neutrinos as  a probe of radically new physics.
Of course, this is a long-shot $\ldots$ but worth trying.

One example of such new physics would be related to the
concept of \emph{emergent symmetries}
\cite{FroggattNielsen1991,Bjorken2001,Laughlin2003,Volovik2003}.
Lorentz invariance, for example, would  not be a fundamental
symmetry but  an emergent phenomenon at low energies.

In order to be specific, we start from an
analogy with quantum phase transitions in fermionic atomic gases or
superconductors and consider the hypothesis
\cite{KV-JETPL2004,KV-IJMPA2005,KV-JETPL2005} that the fermions
of the Standard Model have tiny Lorentz-violating effects due to
Fermi-point splitting (abbreviated FPS and explained below).

If Fermi-point splitting would indeed occur for the quarks and
leptons of the \SM, then
neutrinos may provide a \emph{unique} window to the underlying theory
\cite{K-JETPL2004,K-IJMPA2006,K-PRD2005,K-PRD2006}.
Specifically, there could be new effects in neutrino oscillations,
possibly showing significant \DS{T} and \DS{CP} violation
(and perhaps even \DS{CPT} violation).
The aim of this talk is to sketch some of the potential FPS effects
but we refer, in particular, to the contribution of
M.C. Gonz\'{a}lez-Garc\'{i}a
in these Proceedings for a more general discussion of nonstandard
neutrino oscillations.

The outline of this write-up is as follows.
In Sec.~\ref{FRK-sec:FPS-in-cond-mat}, some background
on condensed matter physics is  given and, in Sec.~\ref{FRK-sec:FPS-hypothesis},
a possible application to elementary particle physics is discussed.
In Sec.~\ref{FRK-sec:Simple-neutrino-model},
a simple but explicit neutrino model with both Fermi-point splittings
and mass differences is introduced.
In Sec.~\ref{FRK-sec:Neutrino-oscillations},
some interesting results on neutrino oscillations from this model are reviewed.
In Sec.~\ref{FRK-sec:Outlook}, concluding remarks are presented.

\section{\label{FRK-sec:FPS-in-cond-mat} Fermi-point splitting
          in atomic and condensed-matter systems}

Ultracold quantum gases of
fermionic atoms (e.g., ${}^6\text{Li}$ at nano-Kelvin temperatures)
are extremely interesting systems, especially as they can have
\emph{tunable} interactions by way of magnetic-field Feshbach
resonances. In the so-called BEC--BCS crossover region
of these systems, a BCS--type condensate has recently been
observed for $s$--wave pairing \cite{Regal-etal2004}.
As usual, BEC stands for Bose--Einstein condensate and
BCS for the superconductivity triumvirate
Bardeen, Cooper, and Schrieffer.

For the BEC--BCS crossover region in systems with
$p$--wave pairing, there is the
prediction \cite{KV-JETPL2004,KV-IJMPA2005} that a
\emph{quantum phase transition}
between a vacuum state with fully-gapped fermionic spectrum and a
vacuum state with topologically protected Fermi points (gap nodes)
occurs. Here, we only give a simple illustration of this new type of
quantum phase transition and refer the reader
to Ref.~\cite{Volovik2006} for an extensive review.

The \BN~Hamiltonian for fermionic quasiparticles in the
axial state of $p$--wave pairing is given by
\beq
H_\text{BN}= \left(
\begin{array}{cc}
|{\bf p}|^2/ (2m)  -q &\;\; c_\perp\,{\bf p}\cdot (\widehat{\bf e}_1+ i\,
\widehat{\bf e}_2)
\\[0mm]
c_\perp\,{\bf p}\cdot (\widehat{\bf e}_1- i\, \widehat{\bf e}_2) &\;\;
-|{\bf p}|^2/ (2m) +q
\end{array} \right),
\label{FRK-BogoliubovNambuH}
\eeq
with
$m$ the mass of the fermionic atom (considered is the direction
of atomic spin, which experiences the Feshbach resonance),
$(\widehat{\bf e}_1,\, \widehat{\bf e}_2,\,\widehat{\bf l}\,)$ an
orthonormal triad, $\widehat{\bf l}$
the direction of the orbital momentum of the pair, $c_\perp$ the
maximum transverse speed, and
$q$ a parameter controlled by the magnetic
field near the Feshbach resonance.

The energy spectrum of this Hamiltonian is readily calculated:
\beq
E_\text{BN}^2 ({\bf p}) = \big(|{\bf p}|^2/(2m)-q \big)^{\!2} +\,
          c_\perp^2\,\big|{\bf p}\times \widehat{\bf l}\:\big|^2 .
\label{FRK-BogoliubovNambuE}
\eeq
Clearly, there are two regimes. For parameter $q<0$, on the one hand,
there is a BEC regime  with mass gap, $E \ne 0$.
For parameter $q>0$, on the other hand, there is a BCS
regime with two Fermi points  in momentum space,
\beq
{\bf b}_1=+p_F \: \widehat{\bf l}\,,\quad
{\bf b}_2=-p_F \: \widehat{\bf l}\,,\quad
p_F\equiv \sqrt{2 m q}\,,
\eeq
at which the energy function vanishes,
$E({\bf p})=0$  for ${\bf p}={\bf b}_a$ with $a=1,2$.

\begin{figure*}[t]
\begin{center}
\includegraphics[width=0.65\textwidth]{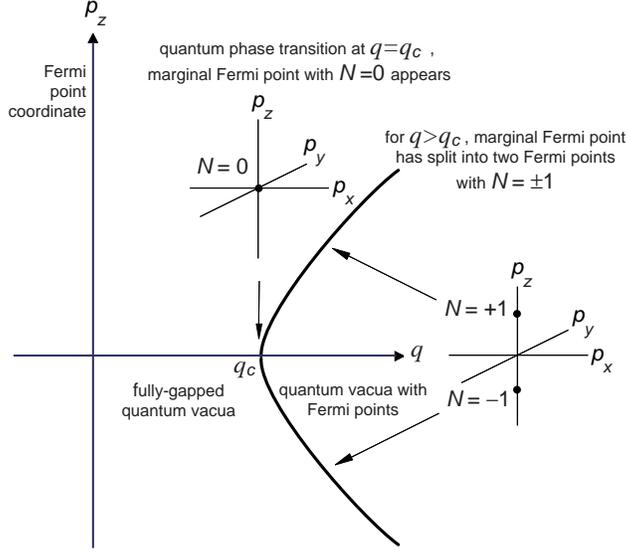}  
\caption{
Quantum phase transition at $q=q_c$ between a quantum vacuum with mass gap and
one with topologically-protected Fermi points (gap nodes).
At $q=q_c$, there appears a marginal Fermi point with topological charge $N=0$
(inset at the top).
For $q > q_c$, the marginal Fermi point has split into two Fermi points
characterized by nonzero topological invariants $N = \pm 1$
(inset on the right). A system
described by Hamiltonian \eqref{FRK-BogoliubovNambuH},
for $\widehat{\bf l}=(0,0,1)$, has critical parameter $q_c=0$.}
\label{FRK-FIG-FPS}
\end{center}
\end{figure*}

There is then a quantum phase transition at $q=0$,
with a mass gap for $q<0$ and a spacelike splitting of Fermi points
($\Delta{\bf b}\equiv{\bf b}_1-{\bf b}_2 \ne 0$)
for $q>0$; see Fig.~\ref{FRK-FIG-FPS}.

This example also clarifies the
concept of emergent relativity mentioned in the Introduction.
Consider momenta close to one of the two Fermi points,
for example, ${\bf p}={\bf b}_1+{\bf k}$ with $|{\bf k}| \ll p_F$.
Then, the energy \eqref{FRK-BogoliubovNambuE} becomes
\beq
E_\text{BN}^2 \sim
(p_F/m)^2\:k_\parallel^2 + c_\perp^2\:k_\perp^2 \sim \, \widetilde{c}^2 \,
\big(\, \widetilde{k}_\parallel^2 + \widetilde{k}_\perp^2\, \big)\,,
\label{FRK-mass-shell}
\eeq
after the following rescalings:
\beq
k_\parallel \equiv {\bf k} \cdot \,\widehat{\bf l}
            \equiv (\,\widetilde{c} \,m/p_F) \;\widetilde{k}_\parallel \,,\quad
k_\perp \equiv |\,{\bf k} \times \widehat{\bf l}\,|
            \equiv (\,\widetilde{c}/c_\perp)\; \widetilde{k}_\perp \,,
\eeq
which would be appropriate for a local observer
made of the \emph{same} quasi-particles \cite{Volovik2003,Volovik2006}.
In terms of the rescaled momentum $\widetilde{\bf k}$, relation
\eqref{FRK-mass-shell} corresponds precisely to the mass-shell condition
of a massless relativistic particle.

\section{\label{FRK-sec:FPS-hypothesis} FPS hypothesis
         for elementary particle physics}

Based on the analogy with certain condensed-matter systems
discussed in Sec.~\ref{FRK-sec:FPS-in-cond-mat}, the following
hypothesis has been put forward \cite{KV-JETPL2004,KV-IJMPA2005}:
perhaps the chiral fermions of the Standard Model also have
Fermi-point splitting (FPS).
Specializing to timelike splittings ($\Delta b_0 \ne 0$)
and vanishing Yukawa coupling constants
(i.e., vanishing fermion masses),
their dispersion relations would be given by:
\beq
\big( E_{a,f}({\bf p}) \big)^2  =
\Big(\,c\, |{\bf p}| + b_{0a}^{(f)}\,\Big)^2 \,,
\label{FRK-SMdispLaw-timelike}
\eeq
where $a$ labels the 16 types of massless left-handed Weyl fermions
(including a left-handed antineutrino) and $f$ the $N_\text{fam}$
fermion families (henceforth, we take $N_\text{fam}=3$).
The maximum velocity of the fermions is assumed to be universal
and equal to the velocity of light \emph{in vacuo}, $c$.
Note that we still speak about Fermi-\emph{point} splitting
even though the energy \eqref{FRK-SMdispLaw-timelike} for
$b_{0a}^{(f)}<0$ gives rise to a Fermi \emph{surface}.

One possible FPS pattern is given by the following \emph{factorized}
Ansatz \cite{KV-IJMPA2005}:
\beq
b_{0a}^{(f)} =  Y_a \; \widetilde{b}_{0}^{(f)}\,,
\label{FRK-SMb0pattern}
\eeq
where $Y_a$ are the known hypercharge values of the fermions
and $\widetilde{b}_{0}^{(f)}$ three unknown energy scales.
Independent of the particular FPS
pattern, the dispersion relations of massless left-handed neutrinos
and right-handed antineutrino would be
\beq
\big( E_{\nu_L,f}({\bf p}) \big)^2  =
\Big(\,c\, |{\bf p}| +    b_0^{(f)}\,\Big)^2  \,,\quad
\big( E_{\bar\nu_R,f}({\bf p})\big)^2  =
\Big(\,c\, |{\bf p}| + s\,b_0^{(f)}\,\Big)^2 \,,
\label{FRK-DispLaw-nu}
\eeq
where a value $s = 1$ respects \DS{CPT} and $s=-1$ violates it.

More generally, one may consider for large momentum $|{\bf p}|$:
\beq
 E({\bf p})       \sim
c\, |{\bf p}| \pm b_0 +  m^2 c^4/(2\, c\,|{\bf p}|) +
\text{O} \big(\,1/|{\bf p}|^{2} \,\big)\,.
\label{FRK-DispLaw-b0m}
\eeq
The conclusion is then that the search for possible
FPS effects prefers neutrinos with the highest possible momentum.

At this point, two questions on energy scales arise.
First, what is known experimentally?
The answer is: not very much, apart from the following
upper bounds:
\beq
|b_0^{(e)}|  \lesssim 1\;\mathrm{keV}\,,\quad
\sum_{f=1}^{3}\; m_f \lesssim 100\;\mathrm{eV}\,,
\eeq
from low-energy neutrino physics \cite{DiGrezia-etal2005}
and cosmology, respectively.

Second, what can be said theoretically about the expected
energy scale of FPS?
The answer is: little to be honest, but perhaps
the following speculation may be of some value. For definiteness,
start from a particular emergent-physics scenario
with two energy scales \cite{KV-JETPL2005}:
\begin{itemize}
\item
$E_\text{LV}$ of the fundamental Lorentz-violating fermionic theory;
\item
$E_\text{comp}$ as the compositeness scale of the Standard Model gauge bosons.
\end{itemize}
Taking the LEP values of the gauge coupling constants,
the renormalization-group equations for $N_\text{fam}=3$ give
these numerical values:
\beq
E_\text{comp} \sim 10^{13}\;\mathrm{GeV} \,, \quad
E_\text{LV}\sim 10^{42}\;\mathrm{GeV} \,.
\eeq
The speculation, now, is that perhaps ultrahigh-energy Lorentz
violation \emph{re-enters} at an ultralow energy scale:
\beq
|b_0|\stackrel{?}{\sim} E_\text{comp}^2/E_\text{LV} \sim 10^{-7}\;\mathrm{eV}\,.
\eeq
If correct, this motivates the search for FPS effects at the sub--eV level.

\section{\label{FRK-sec:Simple-neutrino-model} Simple FPS neutrino model}

A general neutrino model with \emph{both} Fermi-point splittings (FPS)
and mass differences (MD) has many mixing angles and complex
Dirac phases to consider (not to mention possible Majorana phases).
In order to get an idea of
potentially new effects, consider a relatively simple FPS--MD
neutrino model \cite{K-PRD2005,K-PRD2006} having
\begin{itemize}
\item
  a standard neutrino mass sector with ``optimistic'' values
  for $\theta_{13}$ and $\delta$;
\item
  a FPS sector with large mixing angles, energy splittings,
  and Dirac phase $\omega$.
 \end{itemize}
Specifically, the mass sector has the following
mass-square-difference ratio, mixing angles, and Dirac phase:
\begin{subequations}\label{FRK-FPS-MD-model}
\beq
R_m \equiv
\fracnew{\Delta m_{21}^2}{\Delta m_{32}^2}
\equiv \fracnew{m_2^2-m_1^2}{m_3^2-m_2^2} =\frac{1}{30}\,,\:\:
\theta_{21} = \theta_{32} = \frac{\pi}{4}\,,\:\:
\sin^2 2 \theta_{13}=\frac{1}{20}\,,\:\: \delta=\frac{\pi}{2}\,,
\label{FRK-FPS-MD-model-masssector}
\eeq
and the FPS sector has energy-difference ratio, mixing angles, and Dirac phase:
\beq
R \equiv \fracnew{\Delta b_0^{(21)}}{\Delta b_0^{(32)}}
         \equiv \fracnew{b_0^{(2)}- b_0^{(1)}}{b_0^{(3)}- b_0^{(2)}}=1\,,\;\;
\chi_{21} = \chi_{32} = \chi_{13} = \omega = \frac{\pi}{4}\,.
\label{FRK-FPS-MD-model-FPSsector}
\eeq
\end{subequations}

For later use, we also define two additional models.
The first additional model is a \emph{pure} FPS
model \cite{K-IJMPA2006} with trimaximal couplings
($\chi_{21}$ $=$ $\chi_{32}$ $=$  $\pi/4$ and
$\chi_{13}$ $=$ $\arctan\sqrt{1/2}\,$),
complex phase $\omega$, and FPS ratio $R$.
At sufficiently high energies,
the model for $R=1$ and  $\omega =\pi/4$
is close to the FPS--MD model mentioned above.

The second additional model is a \emph{pure} MD model with
a mass-square-difference ratio $R_m  =1/30$ and the following more or less
realistic values for the mixing angles and Dirac phase:
$\sin^2 2 \theta_{23}=1$, $\sin^2 2
\theta_{12}=0.8$, $\sin^2 2 \theta_{13}=0.2$, and $\delta=0$.

In the rest of this contribution, these three models will be referred to as
the FPS--MD model, the FPS model, and the MD model, respectively.

\begin{figure*}[t]
\begin{center}
\includegraphics[width=\textwidth]{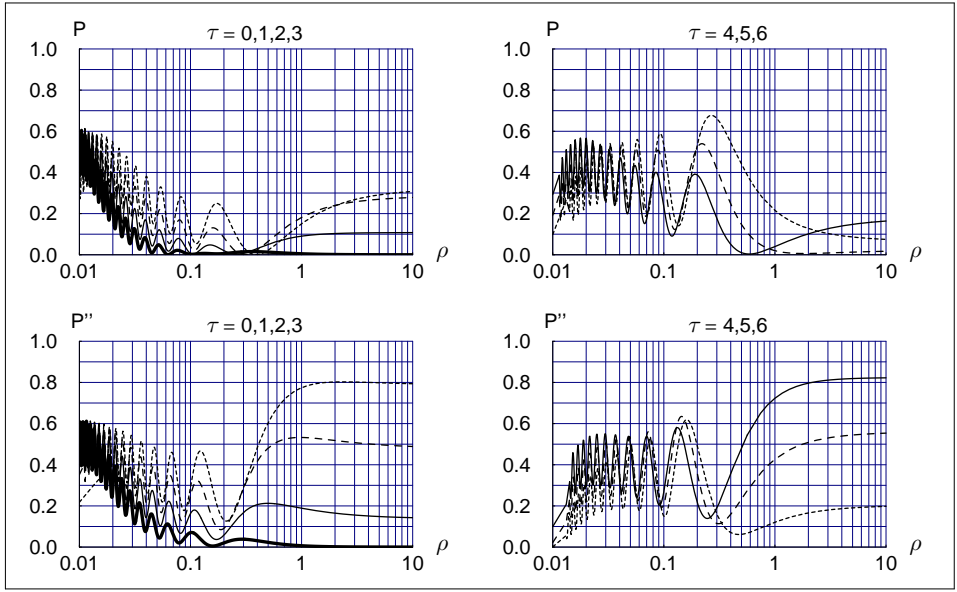}  
\end{center}
\caption{Vacuum probabilities from the FPS--MD model
(\ref{FRK-FPS-MD-model}ab)
as a function of the dimensionless parameters $\rho$ and $\tau$,
defined by Eqs.~(\ref{FRK-rho-tau}ab).
Top panels: $P \equiv P(\nu_\mu\rightarrow \nu_e)$.
Bottom panels: $P^{\prime\prime}
\equiv P(\nu_e\rightarrow \nu_\mu)$.
If \DS{CPT} invariance holds, also $P=P(\overline{\nu}_e\rightarrow
\overline{\nu}_\mu)$ and $P^{\prime\prime}=P(\overline{\nu}_\mu\rightarrow
\overline{\nu}_e)$.
Shown are constant--$\tau$ slices, where the heavy-solid curves
in the two left panels correspond to $\tau=0$ (pure mass-difference model)
and the other thin-solid, long-dashed, and short-dashed curves for positive $\tau$
correspond to $\tau = 1,2,0 \pmod 3$, respectively.}
\label{FRK-FIG-nu-oscill}
\end{figure*}

\section{\label{FRK-sec:Neutrino-oscillations}
         FPS effects in neutrino oscillations}

Consider now a high-energy ($E_\nu \sim c\, |{\bf p}|$)
neutrino beam traveling over a distance $L$.
Neutrino oscillations from the FPS--MD model of
Sec.~\ref{FRK-sec:Simple-neutrino-model}
are then determined by two dimensionless parameters:
\begin{subequations}\label{FRK-rho-tau}
\beq
\rho \equiv
\fracnew{2\,E_\nu\,\hbar c}{L\,|\Delta m^2_{31}|\,c^4} \approx
  0.98 \; \Bigg( \fracnew{E_\nu}{20\;\mathrm{GeV}} \Bigg)
 \Bigg( \fracnew{3000\;\mathrm{km}}{L} \Bigg)\,
 \Bigg( \fracnew{2.7\times 10^{-3}\;\mathrm{eV}^2/c^4}{|\Delta m^2_{31}|}\Bigg),
\label{FRK-rho}
\eeq
\beq
\tau \equiv L\,|\Delta b_0^{(31)}|/(\hbar c) \approx
3.0 \; \Bigg(\fracnew{L}{3000\;\mathrm{km}} \Bigg)
  \Bigg( \fracnew{|\Delta b_0^{(31)}|}{2.0\times 10^{-13}\;\mathrm{eV}} \Bigg).
\label{FRK-tau}
\eeq
\end{subequations}
for numerical values of $L$ and $E_\nu$  appropriate to a neutrino factory
\cite{Geer1997}.
Possible new effects in neutrino oscillations from FPS may occur as
\begin{itemize}
\item
energy dependence of the vacuum mixing angle $\Theta_{13}$
\cite{K-PRD2005};
\item
novel source of \DS{T}, \DS{CP}, and perhaps \DS{CPT} violation
\cite{K-PRD2006};
\item
modified flavor ratios for high-energy cosmic neutrinos
\cite{K-JETPL2004,K-IJMPA2006}.
\end{itemize}
In this contribution, we discuss only the last two effects.

Figure \ref{FRK-FIG-nu-oscill} shows that,
provided the FPS parameter $\Delta b_0^{(31)}$ is large enough
for given baseline $L$, the probabilities of
time-reversed processes can be different by several tens of percents:
$P(\nu_\mu\rightarrow \nu_e)$ $\approx$ $20\,\%$ versus
$P(\nu_e\rightarrow \nu_\mu)$ $\approx$ $80\,\%$
at $\rho \sim 1$ and $\tau \sim 3$, for example.
For the record, standard mass-difference neutrino oscillations
($\tau=0$) give more or less equal probabilities at $\rho \sim 1$:
$P(\nu_\mu\rightarrow \nu_e)$ $\approx$
$P(\nu_e\rightarrow \nu_\mu)$ $\approx$ $0$.
In short, there could be strong \DS{T}--violating
(and \DS{CP}--violating) effects at the high-energy end of
the neutrino spectrum from FPS or other emergent-physics dynamics.

Next, turn to the pure FPS model and also, for comparison, to the
pure MD model, both defined in Sec.~\ref{FRK-sec:Simple-neutrino-model}.
Pion and neutron sources then give the averaged  event ratios
shown in Table~\ref{FRK-table1}, with the clearest difference
between the two models for the case of a neutron
source. In principle, these results may be
relevant to high-energy cosmic neutrinos
but it remains to be seen whether or not
present experiments (e.g., AMANDA and IceCube)
can access this type of information.

\begin{table*}[t]
\caption{\label{FRK-table1}  Averaged  event ratios
$(N_e:N_\mu:N_\tau)$ from pion and neutron sources
for pure Fermi-point-splitting (FPS) and mass-difference (MD)
neutrino models as defined in Sec.~\ref{FRK-sec:Simple-neutrino-model}.
The MD event ratios are taken from Ref.~\cite{Hooper-etal2005}.}
\renewcommand{\tabcolsep}{0.75pc}  
\renewcommand{\arraystretch}{1.1}  
\centering\begin{tabular}{|c|c|c|}
\hline
& $\pi:\text{initial ratios}=(1:2:0)$  & $n: \text{initial ratios}=(1:0:0)$\\
\hline
$\text{FPS} \;(\omega)$ & $( 6 : 7+\cos 2\omega : 5-\cos 2\omega )$
& $( 1 : 1 : 1 )$ \\
$\text{FPS} \;(\pi/4)$  & $(0.33 : 0.39 : 0.28)$  &  $(0.33 : 0.33 : 0.33)$ \\
$\text{MD}$ & $(0.36 : 0.33 : 0.31)$ & $(0.56 : 0.26 : 0.18)$\\
\hline
\end{tabular}
\end{table*}

\vspace{.5\baselineskip}
\section{\label{FRK-sec:Outlook} Outlook}

From a phenomenological perspective,
the Fermi-point-splitting (FPS) hypothesis suggests the
following three research directions:
\begin{itemize}
\item
the possible energy dependence of the vacuum mixing angle $\Theta_{13}$ from
FPS, which can be tested by neutrino experiments at a superbeam
or neutrino factory;
\item
the possibility of a new source of leptonic \DS{CP} violation, which
impacts on the physics of the early universe (e.g., the creation
of baryon and lepton number);
\item
the possible modification of the propagation of high-energy cosmic
neutrinos by FPS effects, which may be of relevance to
present and future neutrino telescopes.
\end{itemize}
From a more theoretical perspective, the outstanding issues are:
\begin{itemize}
\item
the precise nature of the conjectured re-entrance mechanism
of Lorentz violation at ultralow energy from Lorentz violation at ultrahigh
energy (condensed-matter physics can perhaps provide some guidance);
\item
the explanation of the large hierarchies of basic scales
(e.g., for mass or FPS).
\end{itemize}
But apart from these theoretical ideas, experiment may, of course,
suggest entirely different directions  $\ldots$

\vspace*{-1.5\baselineskip}
\section*{Acknowledgments}

{\small
It is a pleasure to thank the EPNT06 organizers and the
Department of Theoretical Physics of Uppsala University for
hospitality and G.E. Volovik for a most stimulating collaboration.
}

\vspace*{-2\baselineskip}

\setcounter{section}{0} \setcounter{subsection}{0} \setcounter{figure}{0}
\setcounter{table}{0}
\newpage
\end{document}